\documentclass{Interspeech2024}

\interspeechcameraready

\title{Navigating the United States Legislative Landscape on Voice Privacy: Existing Laws, Proposed Bills, Protection for Children, and Synthetic Data for AI}

\name{Satwik}{Dutta}
\name{John}{H.L. Hansen}

\address{
  Center for Robust Speech Systems (CRSS), The University of Texas at Dallas, USA}
\email{satwik.dutta@utdallas.edu,john.hansen@utdallas.edu}


\begin{document}

\maketitle

\begin{abstract}
    Privacy is a hot topic for policymakers across the globe, including the United States. Evolving advances in AI and emerging concerns about the misuse of personal data have pushed policymakers to draft legislation on trustworthy AI and privacy protection for its citizens. This paper presents the state of the privacy legislation at the U.S. Congress and outlines how voice data is considered as part of the legislation definition. This paper also reviews additional privacy protection for children. This paper presents a holistic review of enacted and proposed privacy laws, and consideration for voice data, including guidelines for processing children's data, in those laws across the fifty U.S. states. As a groundbreaking alternative to actual human data, ethically generated synthetic data allows much flexibility to keep AI innovation in progress. Given the consideration of synthetic data in AI legislation by policymakers to be relatively new, as compared to that of privacy laws, this paper reviews regulatory considerations for synthetic data.
\end{abstract}

\section{Introduction}

Human voice or speech contains very personal information about a speaker and therefore, it is important to safeguard voice or audio recordings of a speaker from misuse. Guidelines on the collection, storage, and use of any individual's personal data, as collected by any business (such as a company, operator, or service provider), need to comply with the privacy policies as set forward by the local, state, and federal agencies and government. Given numerous recent instances of violation of consumer privacy as well as rapidly evolving Artificial Intelligence (AI) technology which is now available to scammers, lawmakers across U.S. and the world have placed the topic of data privacy on the center stage. Countries and diplomatic organizations across the globe are drafting and implementing AI governance legislation and policies. The thrust for legislation on AI in the United States (U.S.) has been both from the executive branch i.e. the office of the U.S. President or the White House, and the legislative branch i.e. U.S. Congress - collectively by the Senate and the House. Privacy has been a key focus for lawmakers in drafting AI legislation. However, many policymakers agree that the legislation on AI and data privacy would be interlinked. 

In 2022, the White House Office of Science and Technology Policy released the \textit{Blueprint for the AI Bill of Rights} \cite{ai-bill} with outlining principles and good practices to design, use, and deploy AI systems for protecting civil rights, civil liberties, and privacy of the citizens. This was followed by the \textit{Executive Order on Safe, Secure, and Trustworthy AI} \cite{eo-biden} by the President in 2023. This extensive executive order covered safety and security of AI technology, promotion of innovation and competition, support for workforce, protection of rights and privacy, including actions for several federal agencies such as the National Institute of Standards and Technology or NIST. 
On May 15, 2024, a \textit{Roadmap for Artificial Intelligence Policy in the United States Senate} \cite{ai-roadmap} was released  by the Bipartisan Senate Artificial Intelligence (AI) Working Group. This roadmap highlights various policy priorities including funding for AI innovation, enforcement of existing laws for AI, impact of AI on workforce, enhancing national security, addressing challenges posed by deepfakes, and support for higher education research and development on AI. This roadmap also prioritizes policies on establishing a strong comprehensive federal data privacy framework.    

\begin{figure*}[h]
    \centering
    \includegraphics[width=1.0\linewidth]{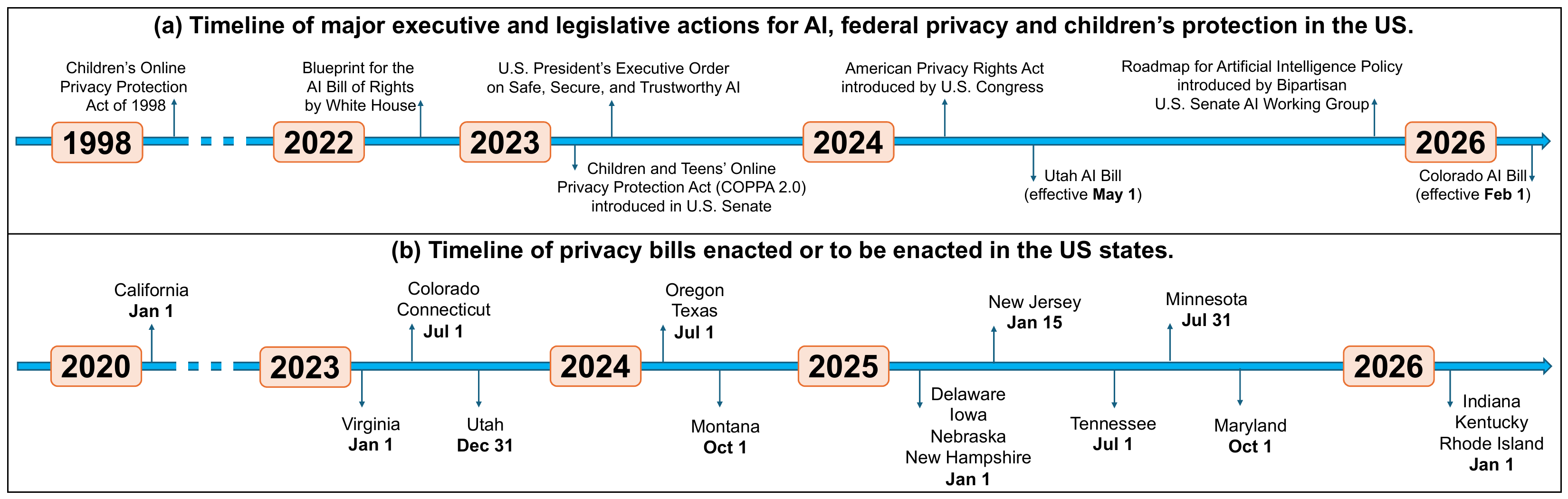}
	\caption{(a) Timeline of major executive and legislative actions for AI, federal privacy and children’s protection in the US (as of Jun 24, 2024). (b) Timeline of privacy bills enacted or to be enacted in the US states (as of Jun 24, 2024).}
	\label{fig:timeline}
\end{figure*}

Motivated by many recent U.S. legislation on privacy protection, this paper aims to give an overview and current state of both the federal and state privacy policies across the U.S. A holistic review of these privacy legislation also highlights how voice fits in the legislative definition. Children's data being sensitive \cite{dutta22_interspeech,dutta2022activity,dutta24_odyssey}, a review of protection for children's privacy legislation is also presented in this paper. As compared to privacy laws, AI legislation is new, and many U.S. states are considering additional AI legislation on top of the privacy laws. Some of these AI legislation also consider defining synthetic data and voice generation. An illustrative timeline of major legislative actions for AI and privacy is shown in Fig.\ref{fig:timeline}.
This paper is structured as follows: a newly proposed national privacy legislation for the U.S. is discussed in Sec.\ref{sec:apra}, followed by privacy protection guidelines for children in Sec.\ref{sec:child_privacy}. This is followed by a holistic review of the privacy laws across all the U.S. states in Sec.\ref{sec:state-privacy} and laws on synthetic data for AI in Sec.\ref{sec:syn-data}. Finally, we conclude this paper in Sec.\ref{sec:con}. 

\section{American Privacy Rights Act of 2024}
\label{sec:apra}

The \textit{American Privacy Rights Act of 2024} (APRA) \cite{arpa,arpa-ann} was proposed in April 2024 by two prominent members of the U.S. Congress who serve as Chairs of the House Committee on Energy and Commerce and Senate Committee on Commerce, Science and Transportation. As the lawmakers mentioned, this proposed legislation is ``bipartisan'' (support from lawmakers from both political parties) and  ``bicameral" (support from lawmakers from both chambers of the U.S. Congress). The primary goal of APRA is to establish a nationwide comprehensive data privacy and security standard for all U.S. citizens including children. APRA includes several actions such as (1) guidelines on prohibitions on consumer personal data use, including sensitive and biometric data of consumers, (2) consumer rights, including access, correction, and deletion, on covered data (covered data is any data which identifies or is linked or reasonably linkable to an individual), (3) policies for design by privacy and transparency, (4) opt-out privilege of consumers, and (5) enforcement of internal data security and privacy guardrails by businesses. 

This legislation primarily includes all businesses subject to the U.S. Federal Trade Commission's (FTC) authority, common carriers, and non-profits, except small businesses ($<$USD 40 million in annual revenue, $<$200,000 consumers). This also includes additional obligations for high-impact social media companies, data holders, and data brokers. After its original draft was released in April, APRA has been discussed and edited twice with an expected markup (lawmakers offer and vote on amendments) scheduled for June 27, 2024, by the House Committee on Energy and Commerce. However, there are still multiple concerns with this legislation. 

APRA considers ``\textbf{voice prints}'' as biometric information, which is defined as any data that could directly or indirectly help to identify an individual and that could be generated through unique characteristics of an individual like biological, physical, or physiological. Biometric information is considered as ``sensitive covered data''. There are guidelines regarding the collection, processing, retention, or transfer of biometric information, and the need for consent from consumers. An ``\textbf{audio recording}" or any data derived from such a recording is \textbf{excluded} from the definition of biometric information unless it could be used to identify an individual. Private communications of an individual such as ``voicemails'', ``voice or video communications'', or any related information regarding its transmission and private ``audio recordings'' are also considered as ``sensitive covered data''. Additional protections are outlined regarding the transfer of sensitive covered data. 

\section{Children's Privacy in the US}
\label{sec:child_privacy}

\subsection{1998 Children's Online Privacy Protection Act}

The \textit{Children's Online Privacy Protection Act of 1998} \cite{coppa}, COPPA, was enacted to prohibit unfair collection and use of personal information of children under the age of 13 on the web. In 2013, COPPA was amended to extend the definition of ``website or online service directed to children'', to expand the definition of ``personal information'' to include \textbf{audio file} where such file contains a child's ``voice'' (including photo, video), and acceptable methods for verifiable parental consent.

\subsection{2024 Children \& Teens’ Online Privacy Protection Act}

The Children and Teens’ Online Privacy Protection Act of 2024 \cite{coppa2}, COPPA 2.0, was recently proposed in the U.S. Congress as an update to COPPA. This particular legislation has gained significant momentum since Feb 2024, both at the U.S. Senate and House. It has also been discussed in the House Committee on Energy and Commerce along with other bills such as KOSA (Kids Online Safety Act). COPPA 2.0 would also extend protection to teens between 12 to 17 years in age. Several updates have been proposed in the new COPPA 2.0 legislation, including extension of the definition of personal information, website/online service providers, consent, data retention, advertising, and use for educational technology. 

Particularly considering voice, the definition of personal information has been modified for one item and a new item added: (1) ``\textit{an \textbf{audio} file where such file contains a specific child's or teen's \textbf{voice}}'', and (2) ``\textit{information generated from the measurement or technological processing of an individual's biological, physical, or physiological characteristics that is used to identify an individual, including \textbf{voice prints}}''. COPPA 2.0 also adds a specific \textbf{exclusion section for audio files}. Audio files are excluded from being considered as personal information if the service or operator: (1)  does not request for personal identifiable information, (2) clearly states the collection, use, and deletion policy in its privacy policy, (3) only uses the audio file for the intent or task for which it was collected, (4) maintains the audio file to perform the intent or task, and deletes the audio file immediately without any other use before deletion. These guidelines suggest that \textbf{the audio files could only be used for providing a service, and not for any innovation of the product or service}. There is currently no legislative text on de-identification or use of de-identified data. As of May 2024, COPPA 2.0 has been added as a section under APRA. The primary goal for COPPA 2.0 is to bring online data privacy protection for children and teens to the 21st century. 

\section{State-level Privacy Regulations in the US}
\label{sec:state-privacy}

\begin{figure*}[h]
    \centering
    \includegraphics[angle=90,width=0.9\linewidth]{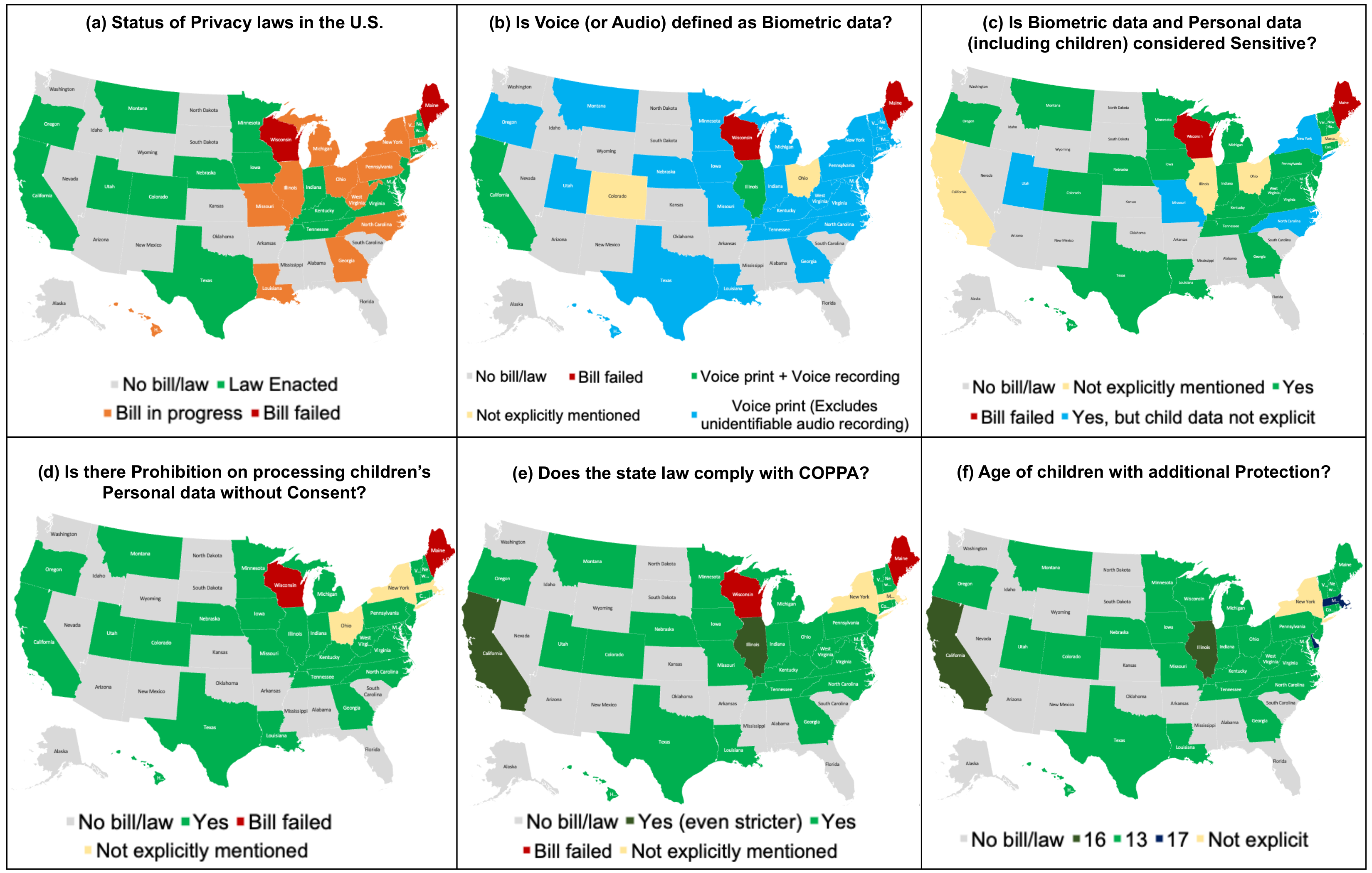}
	\caption{(a) Status of Privacy laws in the U.S. (b) Is Voice (or Audio) defined as Biometric data? (c) Is Biometric data and Personal data (including children) considered Sensitive? (d) Is there a Prohibition on processing children’s Personal data without Consent? (e) Does the state law comply with COPPA? (f) Age of children with additional Protection?}
	\label{fig:state}
\end{figure*}

Out of all the 50 U.S. states, California was the first state to enact a strict consumer privacy law in 2020 - the \textit{California Consumer Privacy Act of 2018} (CCPA) \cite{ccpa}. Additional privacy protections were added to CCPA in 2020 and amended by the \textit{California Privacy Rights Act of 2020}. Founded in 2020 by CCPA, the California Privacy Protection Agency began updating existing laws and adopting new legislation in 2022. The action on state privacy bills was followed by other states such as: Virginia, Colorado, Connecticut, Utah, and many more. 19 U.S. states have already enacted new privacy legislation. Legislation on privacy has been proposed and is in progress in 13 states, while the action has already failed in 2 states. To date, no privacy bill has been proposed in the remaining 16 U.S. states. An illustrative map of the current status of privacy laws across all U.S. states is shown in Fig.\ref{fig:state}(a).  

Particularly for legislative text on ``\textbf{voice}'', based on our review,  most states consider ``\textbf{voice prints}'' as biometric information similar to ARPA. While only California \cite{ccpa} and Illinois \cite{illi} add ``\textbf{voice recording}" in the definition of biometric information, most states exclude ``\textbf{audio recording}" or any data derived from such a recording as biometric information unless it could be used to identify an individual.  Delaware \cite{dela} and Pennsylvania \cite{penn} also add another specific exclusion\footnote{Information captured and converted to a mathematical representation, including a numeric string or similar configuration, that cannot be used to recreate data generated by automatic measurement of an individual's biological patterns or characteristics used to identify the specific individual} in their ``biometric information'' definition which could refer to raw audio when converted to an array or acoustic features. Most states also define a broader ``sensitive data'' which includes personal and biometric data (or information). The legislative text for Ohio \cite{ohio} is very different from all other states, not explicitly defining biometric or sensitive data or including voice or audio recordings. Infographics showing which states include voice in the definition of biometric information and which states add a broader definition of sensitive data (including biometric and personal) are shown in Fig.\ref{fig:state}(b,c).  

Many states also add additional protection for children's privacy by prohibiting the processing of sensitive (or personal) data of children with verifiable consent as shown in Fig.\ref{fig:state}(d). While most states comply with other state and federal regulations, some explicitly mention and comply with COPPA, as shown in Fig.\ref{fig:state}(e). It should be noted that California \cite{ccpa} and Illinois \cite{illi} have even stricter child privacy regulations as compared to the enacted version of COPPA. While most states consider the protection of child privacy rights below 13 years, few keep the limit up to 16 or 17 years, as shown in Fig.\ref{fig:state}(f).

\section{Regulations on Synthetic Data for AI}
\label{sec:syn-data}

Utah \cite{utah-ai} is the first U.S. state to enact a comprehensive AI governance law that went into effect on May 1, 2024. Similar to its privacy law \cite{utah-p}, ``\textbf{voice}'' is defined in the AI legislative text. In this legislation, ``synthetic data'' is defined as ``\textit{data that has been generated by computer algorithms or statistical models and does not contain personal data}'', therefore considering synthetic data as de-identified data. The AI legislation in Utah is followed by Colorado \cite{colorado-ai}, which will be effective on February 1, 2026. This legislation majorly focuses on the risks of AI systems related to discrimination, including many other arguments. ``\textbf{Voice}'' or ``\textbf{synthetic data}'' is not defined in the legislation, but this legislation does not consider conversational AI technology a high risk unless it generates content that is discriminatory or harmful. AI legislation has passed in only these two U.S. states, and in progress in several states: California, Illinois, New York, Louisiana, Massachusetts, Ohio, and Oklahoma. 

The proposed California AI Transparency Act \cite{cali-ai} would require any person that creates, codes, or otherwise produces a generative AI system, which is publicly available and has over 1,000,000 monthly visitors, to include a latent disclosure (permanent, to the extent it is technically feasible) in AI-generated digital content (synthetic data) including ``audio''. Illinois \cite{illi-ai} Consumer Fraud-AI legislation proposes for requirement of disclosure on synthetic media in advertising. However, the definition also includes ``human voice'' \textit{created, reproduced, or modified by generative AI or a software algorithm to produce or reproduce a human voice}. Similar to Illinois, the AI legislation in New York \cite{ny-ai} requires advertisements to disclose the use of a synthetic performer (synthetic data), but does not explicitly mention ``voice''. 
Louisiana's proposed AI bill \cite{lou-ai} primarily revolves around guidelines for AI foundation models, requiring every publicly available (made available by any person in the state) foundation model and its use to be officially registered with the state, and await for further guidelines from the state. ``Audio'' is explicitly defined as one of the ``AI-generated content'' (synthetic data) either created or modified by a ``generative artificial intelligence system'' in the proposed AI legislation for the Commonwealth of Massachusetts \cite{mass-ai}. A mandatory disclosure is required for all generative AI systems, otherwise punishable, with notice and metadata information of the AI-generated content. Any person using such a Generative-AI system to generate or re-purpose AI-generated content would also be prohibited from removing the disclosure information. The proposed AI legislation in Ohio \cite{ohio-ai} provides guidelines on AI-generated (or synthetic) products and the prohibition of identity fraud using a replica of a person. ``Replica of a person's persona'' (or replica) is defined as a customized version of an individual's ``voice'' (and other factors), that appears to be the individual's authentic persona. The replica could be partially or fully generated by AI. The proposed AI legislation in Oklahoma \cite{ok-ai} gives its citizens the right to consent to any ``\textit{derivative media that is generated by an artificial intelligence engine and uses audio recordings of the citizen's voice or images of him or her to recreate the citizen's likeness}''.

\section{Conclusion}
\label{sec:con}

Innovation in AI is necessary, but not at the cost of privacy. The speech technology community has led several efforts on voice privacy including The Voice Privacy Challenge \cite{tomashenko2022voiceprivacy} and Symposium on Security and Privacy in Speech Communication. It is important to communicate such efforts on voice privacy to the public and policymakers. Businesses have an obligation to disseminate whether they follow privacy-preserving technology development, and also how they collect, use, and maintain consumer data. Such actions would be valuable for speech technology research and development to sustain, as the world navigates regulations on privacy and trustworthy AI. On the legislative actions, momentum on both privacy and AI bills is rapid. Apart from many regulations, California and Illinois have proposed Consumer Privacy Funds to educate the public, including children in the area of online privacy. In the future, it would be interesting to navigate and compare how the legislative landscape in the U.S. differs from other countries across the globe such as the GDPR and the EU AI Act.   

\section{Acknowledgements}
This work is sponsored by the Quad Fellowship and McDermott Graduate Fellowship (Dutta), and NSF awards \#1918032, \#2234916, and \#2341384 (Hansen). Partial work was done for the URA Science Policy Competition (SPARC) and an AI Legislation Sprint by the Federation of American Scientists. 

\bibliographystyle{IEEEtran}
\bibliography{mybib}

\end{document}